\documentclass[12pt]{article}

\usepackage[left=1.2in,right=1.2in,top=1.2in,bottom=1.2in]{geometry}
\usepackage{hyperref}
\usepackage{graphicx}
\usepackage{amssymb}
\usepackage{amsbsy}
\usepackage{amsmath}
\usepackage{hyperref}
\usepackage{xcolor}
\hypersetup{
    colorlinks,
    linkcolor={red!50!black},
    citecolor={blue!50!black},
    urlcolor={blue!80!black}
}

\newcommand{\rf}[1]{(\ref{#1})}
\newcommand{\beq}{\begin{equation}}
\newcommand{\eeq}{\end{equation}}
\newcommand{\be}{\begin{equation}}
\newcommand{\ee}{\end{equation}}
\newcommand{\bea}{\begin{eqnarray}}
\newcommand{\eea}{\end{eqnarray}}

\newcommand{\non}{\nonumber \\*}

\newcommand{\bv}{\begin{verbatim}}
\newcommand{\ev}{\end{verbatim}}

\def\XXint#1#2#3{{\setbox0=\hbox{$#1{#2#3}{\int}$}
     \vcenter{\hbox{$#2#3$}}\kern-.5\wd0}}

\newcommand{\e}{\,\mbox{e}}
\renewcommand{\d}{{\rm d}}

\newcommand{\sqL}{\sqrt{\Lambda} }
\newcommand{\sqZ}{\sqrt{Z}}
\newcommand{\eps}{\epsilon}
\renewcommand{\a}{\alpha}
\newcommand{\om}{\omega}
\newcommand{\oh}{\frac{1}{2}}
\newcommand{\Vc}{V_{cnt}}
\newcommand{\Vp}{V'_{cnt}}

\newcommand{\tit}{\tilde{t}}
\newcommand{\tR}{\tilde{R}}

\newcommand{\tr}{\mathrm{tr}}
\newcommand{\LA}{\left\langle}
\newcommand{\RA}{\right\rangle}
\newcommand{\prt}{\partial}

\newcommand{\tV}{\tilde{V}}

\newcommand{\cM}{{\cal M}}

\newcommand{\tM}{\tilde{M}}

\newcommand{\half}{\textstyle{\frac 12}}
\newcommand{\wt}{\widetilde}
\newcommand{\tT}{\widetilde{T}}
\newcommand{\mth}{m^{{\rm th}}}

\newcommand{\plus}{\!+\!}
\newcommand{\minus}{\!-\!}

\def\fun#1#2{\lower3.6pt\vbox{\baselineskip0pt\lineskip.9pt
\ialign{$\mathsurround=0pt#1\hfil##\hfil$\crcr#2\crcr\sim\crcr}}}

\begin{document}

\begin{center}
\vspace{24pt}
{ \large \bf 
Perturbed generalized multicritical one-matrix models
}

\vspace{30pt}

{\sl J. Ambj\o rn}$\,^{a,b}$,  {\sl L. Chekhov}$\,^{a,c,e}$ and {\sl Y. Makeenko}$\,^{a,d}$

\vspace{24pt}
{\footnotesize

$^a$~The Niels Bohr Institute, Copenhagen University\\
Blegdamsvej 17, DK-2100 Copenhagen, Denmark.\\

\vspace{10pt}

$^b$~IMAPP, Radboud University, \\
Heyendaalseweg 135, 6525 AJ, Nijmegen, The Netherlands\\
{ email: ambjorn@nbi.dk}\\

\vspace{10pt}

$^c$~Steklov Mathematical
Institute and  Laboratoire Poncelet \\
Gubkina 8, 119991
Moscow, Russia,\\
{ email: chekhov@mi.ras.ru.}\\

\vspace{10pt}

$^d$~Institute of  Theoretical and Experimental Physics\\
B. Cheremushkinskaya 25, 117218 Moscow, Russia.\\
{ email: makeenko@nbi.dk}\\

\vspace{12pt}

$^e$~Michigan State University, 
 East Lansing, USA

}
\end{center}
\vspace{48pt}

\begin{center}
{\bf Abstract}
\end{center}
We study perturbations around the generalized   Kazakov multicritical one-matrix model.
The multicritical matrix model has a potential where the coefficients of $z^n$ 
only fall off as a power $1/n^{s+1}$. This implies that the potential and its 
derivatives have a cut along  the real axis, leading to technical problems when 
one performs perturbations  away from the generalized Kazakov model. 
Nevertheless it is possible to relate the perturbed partition function to 
the tau-function of a KdV hierarchy and solve the model by a genus expansion 
in the double scaling limit.

\newpage

\section{Introduction}

The one-matrix model has been used 
to describe two-dimensional quantum gravity coupled to certain 
conformal field theories. More precisely, by fine-tuning the potential 
used in the matrix integrals to the so-called Kazakov potentials \cite{kazakov2} one
could obtain scaling limits which describe rational $(p,q) =(2,2m \minus 1)$ conformal 
field theories coupled to two-dimensional quantum gravity, $m=2,3,\ldots$ \cite{staudacher}. 
One can think of the matrix models as a lattice regularization of an underlying continuum 
theory of two-dimensional quantum gravity coupled to matter fields, where
the fine-tuning of the potential corresponds to taking the lattice spacing to 
zero\footnote{Historically, this interpretation came from attempts to provide a
non-perturbative regularization of the bosonic string using the Polyakov formulation
\cite{history}. The quantum gravity interpretation of ensembles of triangulations 
was later extended to higher dimensions \cite{higherd}, most successfully maybe 
in the context of the so-called causal dynamical triangulations (see \cite{physrep} for a review and 
\cite{higherCDT} for the original articles). However, contrary to 
the 2d situation, it is still unclear if these higher dimensionall theories really 
define a theory of quantum gravity.} 

It is possible to fine-tune a potential to the Kazakov potential 
in several ways. The various fine-tunings have the interpretation as corresponding to
continuum theories deformed away from the pure conformal field theory by the 
primary operators of the $(2,2m\minus 1)$ theory (for a review see \cite{zinn}). There are $m\minus 1$ primary operators
associated with these deformations and in this sense one has a nice Wilsonian 
picture: the matrix model can be formulated with many (potentially infinitely many)
coupling constants and by approaching the $m$'th multicritical point there 
exists $m\minus 1$ relevant coupling constants which define the continuum limit,
while the rest are irrelevant for this limit. Thus the critical surface associated with
the $m$'th multicritical point will constitute a hypersurface of co-dimension $m\minus 1$ 
in the infinte dimensional coupling constant space and the continuum coupling 
constants related to the primary operators are fixed by the way one approaches the 
critical surface.\footnote{One aspect which is missing when comparing to the standard 
Wilsonian picture is the {\it divergent correlation length} of two-point functions.
This correlation length is of course the main organizing quantity in the Wilsonian 
picture, and it is an interesting question how to introduce it in the theory where 
quantum gravity is involved and where there is no fixed background geometry. There 
is good evidence that such a correlation length exist for 2d quantum gravity 
coupled to matter, but much remains to be understood \cite{dt-matter}.}

Another aspect of the continuum limit of the one-matrix models is that the singular part 
of the matrix integral can be identified 
with the tau-function of a 2-reduced KdV hierarchy \cite{KdV} (see \cite{Rev} for a brief review).
While the detailed relationship between the matrix model 
coupling constants and continuum coupling constants of the deformed conformal field theories coupled to gravity 
is somewhat complicated,  there exists a relatively simple matrix model prescription which allows us to 
express the parameters of the KdV hierarchy in terms of the matrix model coupling constants.

In this article we will describe how some of these well known results for the ordinary one-matrix models
are valid also for the generalized one-matrix model considered in \cite{abm}, where the potential
has a long tail of the Taylor 
expansion, leading to a cut of the potential and its derivative at the real axis,
and reduces to polynomials at Kazakov's multicritical points.
The rest of the article is organized as follows: in Sec.\ \ref{oldstuff} we shortly describe the relevant results 
for the ordinary one-matrix model. In Sec.\ \ref{newstuff} we show how the KdV structure is somewhat 
different for the generalized one-matrix models and Sec.\ \ref{GD} discusses the modifications of the 
string equation encountered for the generalized matrix model.
Finally Sec.\ \ref{discussion}
contains a discussion of the results obtained.

\section{Ordinary multicritical matrix models}\label{oldstuff}

\subsection{The loop equation}

We consider the $N\times N$ Hermitian matrix model
\beq\label{ja2}
Z = \int d M \, \e^{-N \tr V(M)},
\eeq
with an even potential which is a polynomial 
\beq\label{ja3}
V(x) = \frac{1}{g} \tilde{V}(x), \quad 
\tilde{V}(x)=\sum_{n=1}^m t_{2n} x^{2n}.
\eeq
In the large-$N$ limit there is a one-cut solution, where the eigenvalues of $M$ 
condense in an interval $[-a,a]$. The so-called  resolvent or disk amplitude
\beq\label{ja3W}
w(z) = \frac{1}{N} \LA \tr \;\frac{1}{z-M} \RA  = \int_{-a}^a \d x \, \frac{\rho(x)}{z-x}
\eeq
is an analytic function of $z$ outside the cut. 

The large-$N$ solution for $w(z)$ is 
\be
w(z)=\int_0^a \frac{\d x}{\pi}\frac{xV'(x)}{(z^2-x^2)} 
\frac{\sqrt{z^2-a^2}}{\sqrt{a^2-x^2}},
\label{Weven}
\ee
where the condition $w(z) \to 1/z$ for $|z| \to \infty$ implies
\be
g(a^2)=\int_0^a \frac{\d x}{\pi}\frac{x\tV'(x)}{\sqrt{a^2-x^2}} \;=\;
 \sum_{n=1}^m  \frac{t_{2n} \,a^{2n}}{B(n,\frac 12)},
 \quad\quad B(x,y)=\frac{\Gamma(x)\Gamma(y)}{\Gamma(x+y)}.
\label{bbcc}
\ee
 This fixes $g(a^2)$ as a polynomial of $a^2$ for a  polynomial potential. 
 For further reference we note that one can rewrite \rf{Weven} as (see  \cite{abm} for details)
 \beq\label{jaa4}
 g w(z) = \int_0^{a^2}  d \tilde{a}^2\;\frac{g'(\tilde{a}^2)}{\sqrt{z^2 -\tilde{a}^2}} 
 =\int_0^g \frac{d\tilde{g}}{\sqrt{z^2 - a^2(\tilde{g})}},
 \eeq
 which implies
 \beq\label{jab4}
 \frac{ d \;(gw(z))}{dg} = \frac{1}{\sqrt{z^2-a^2}}.
 \eeq
 This is the simplest manifistation of the universality of matrix model correlators discovered 
 in \cite{am,ajm,ackm}, and it will play an important role below.  
 
 A so-called  $m^{\rm th}$ multicritical point
 of this matrix model is a point where 
 \beq\label{ja4}
 \frac{\d g(A)}{\d A}\Big|_{A=a^2_c} = \cdots = 
\frac{\d^{m-1} g(A)}{\d A^{m-1}}\Big|_{A=a^2_c} =0, \quad 
\frac{\d^{m} g(A)}{\d A^{m}}\Big|_{A=a^2_c} \neq 0.
 \eeq
 In order to satisfy this  requirement an even potential $\tV (x)$ has to be at least
 of order $2m$. If we restrict ourselves to potentials of this order, $g(a^2)$ is fixed 
 to be of the form
 \beq\label{ja5}
 g(a^2)= g_* - c (a_c^2 -a^2)^m, \quad g_* = c \,a_c^{2m}, \quad  c = \frac{1}{4m a_c^{2m-2}}.
 \eeq
The value $a_c^2 > 0$ can be chosen arbitrary, after which the coefficients $t_n$ are completely fixed
in the case where the polynomial is of order $2m$.  We denote this potential {\it the Kazakov potential }
$V^{(K)}(x)$. 

If we allow higher order polynomials, the critical point 
will become a critical surface in the coupling constant space. This critical surface is conveniently characterized 
by introducing the so-called moments $M_k$, defined by 
 \beq\label{jh1}
 M_k(g,t;a) = \oh \oint_C \frac{\d z}{2\pi i} \; \frac{z V'(z)}{(z^2-a^2)^{k+1/2}} = \frac{\Gamma(\oh)}{\Gamma(k-\oh)}
 \int_{0}^1  \frac{\d x}{\pi }\; \frac{\big(\frac{\prt}{\prt a^2}\big)^{k} ( ax V'(ax))}{(1-x^2)^{1/2}} ,
 \eeq
 where the contour $C$ encircles to cut of $w(z)$. The position of the cut $a$ as a function of the 
 coupling constants $g,t$ is determined by eq.\ \rf{bbcc} which can be written as 
 \beq\label{jh1b} 
 M_0(g,t;a) = 1.
 \eeq
 For a polynomial of order $2n$ only the $n$ moments 
 $M_k$, $k=1,\ldots,n$ can be different from zero and we can express $w_0(z)$ as 
 \beq\label{jh1a}
 w_0(z)- \oh V'(z)  = -M(z^2 \minus a^2) \sqrt{z^2-a^2}, \quad M(z^2\minus a^2) = \sum_{k=1}^n M_k \; (z^2-a^2)^{k-1}.
 \eeq
 
Consider the set of (even) polynomials of arbitrary orders $2n \geq 2m$. 
 The $m$'th multicritcal surface is then  defined by
\beq\label{jh2}
 M_1 = \cdots = M_{m-1}= 0, \quad M_m \neq 0.
 \eeq
Let now $g^*$ and $t^*$  correspond to a specific point on this critical surface. One can approach 
this point in the following way
\beq\label{jh3}
M_k (t,g) = \mu_k \eps^{m-k}, \quad 0 < k \leq m, \quad M_0(t,g) = M_0(t^*,g^*) + \mu_0\eps^m,
\eeq  
where $\mu_m \neq 0$ and $\mu_k = 0$ for $k > m$ while $\eps$ is a parameter scaling to zero.
Eq.\ \rf{bbcc} implies that $\mu_0 = 0$ and $M_0(t,g) = M_0(t^*,g^*) = 1$ is the equation which determines
$a$ as a function of $t,g$.

The {\it scaling limit}\/ is now defined by the additional requirement
\beq\label{jh3b}
z^2 =a_c^2 +\eps Z, \quad a^2 = a_c^2 - \eps \sqL.
\eeq
In this scaling limit\footnote{The factor 2 in the definition of the continuum limit in 
\rf{jh3a} is convention and related to the fact that we have here, for convenience, restricted ourselves
to symmetric potentials. For an asymmetric potential the eigenvalues of the potentials 
would lie in an interval $[-b,a]$ where $b\neq a$ and one would approach the point $a$, i.e.\
one would make a scaling $z = a_c + \eps Z$. Here we choose, to keep the symmetry, the scaling
$z^2 =a_c^2 +\eps Z$, i.e.\ we are approaching both $a$ and $-a$ for $Z \to 0$. Thus we get 
two times the ``real'' continuum limit.} 
we obtain the continuum limit immediately from \rf{jh1a}
\beq\label{jh3a}
 w_0(z)- \oh V'(z)  = -\eps^{m-\oh}  \sum_{k=1}^m \mu_k (Z\plus \sqL)^{k-\oh} :=  
 \eps^{m-\oh} 2\big(W_0(Z) -\oh \Vp(Z)\big).
 \eeq
Since $(Z\plus \sqL)^{n-\oh}/\sqrt{Z}$ has a convergent Laurent expansion for $|Z| > \sqL$ the same is true 
for sum of terms and we define $\Vc(Z)/\sqrt{Z}$ to be the part which contains non-negative powers of $Z$ while 
$W_0(Z)/\sqrt{Z}$ is the part which contains the negative powers of $Z$, starting with $Z^{-2}$. We denote 
$\Vc(Z)$ the {\it continuum potential} and write 
\beq\label{jq1}
\Vc(Z) = \sum_{k=0}^{m} T_k Z^{k+\oh}, \qquad V_{cnt}^{(K)}(Z) = T^{(K)}_0 Z^\oh + T^{(K)}_m Z^{m+\oh}, 
\eeq
where $V_{cnt}^{(K)}(Z)$ is the continuum Kazakov potential where 
\beq\label{jq1a}
\oh T^{(K)}_0 = \Lambda^{\frac{m}{2}}, \quad 
(m\plus \oh)T^{(K)}_m = (-1)^{m-1} \frac{\sqrt{\pi} \Gamma(m+1)}{\Gamma(m+\frac{1}{2})},
\quad T^{(K)}_k=0, ~~k \neq 0,m.
\eeq

The relation between the $T_k$ and the $\mu_k$ follows from  the definition \rf{jh3a}. 
However, it is conveniently shown in the following way, which also reveals  an important property of the $T_k$. 
The moments $M_k$ and correspondingly the $\mu_k$ depend in a complicated way on the coupling
constants $t_k$ because they depend also on the position $a$ of the cut, and $a$ is a complicated
function of the couplings $t_k,g$ .  Let us introduce new modified moments $\tM_k$ which only depend  
 linearly on  $t_k/g$.  Corresponding to $g^*,t^*$ we have a cut $[-a_c,a_c]$ and we  define 
 \beq\label{jh9}
 \tM_k =\oh \oint_C \frac{\d z}{2\pi \,i} \; \frac{z V'(z)}{(z^2-a_c^2)^{k+1/2}} = M_k (g,t;a \to a_c),
 \eeq
where $C$ encircles the cut $[-a_c,a_c]$. It is clear that $\tM_k$ depends linearly on the 
coupling constants $t_k/g$ for fixed $t^*,g^*$, i.e. $a_c$ fixed. 

Let us now approach $t^*,g^*$ according to \rf{jh3}.
Since $w_0(z) \to 1/z$ for $z\to \infty$ we can  write, using $\eps Z = z^2-a_c^2$ and \rf{jh3a}, 
\bea\label{jq2}
\tM_k &=& \oh \oint_{C_1} \frac{\d z}{2\pi \,i} \; \frac{z \tV'(z)-2w_0(z)}{(z^2-a_c^2)^{k+1/2}} =  \eps^{m-k} 
\oint_{C_2} \frac{\d Z}{2\pi \,i} \; \frac{\Vp(Z)- 2W_0(Z))}{Z^{k+1/2}} \\
&=&  \oint_{C_3} \frac{\d Z}{2\pi \,i} \; \frac{\Vp(Z)}{Z^{k+\oh}} =   \eps^{m-k} (k\plus \oh) T_k.
\eea
Starting out with the contour $C$ from \rf{jh9}, the integrand in the first line-integral in \rf{jq2} has 
two  cuts on the real axis, $[a,a_c]$ and $[-a_c,-a]$. $C_1$ is $C$ contracted to curves encircling these cuts.
By mapping to the $Z$-plane, $C_1$ is mapped to two curves encircling the cut $[-\sqL,0]$ of  $W_0(Z)/\sqZ$
(and the pole $T_0/Z$ of $\Vp(Z)/\sqZ$). $C_2$ denotes one of these curves. The line-integral involving 
 $W_0(Z)/\sqZ$ is zero  since $C_2$ can be deformed to infinity and $W_0(Z)/\sqZ$ has a convergent
 power expansion in $1/Z$ starting with $1/Z^2$. For the remaining term $\Vp(Z)/\sqZ$ we can deform 
 $C_2$ to any curve $C_3$ encircling 0.

We thus see that the approach \rf{jh10} to $t^*,g^*$ alternatively can be formulated as 
\beq\label{jh10}
\tM_k =   (k\plus \oh) T_k \eps^{m-k}, \quad 0 < k \leq m, \qquad \tM_0 = \tM_0^* + \frac{1}{2}\,T_0 \eps^m,
\eeq
where $\tM^*_0 = \tM_0(g^*,t_k^*) ~(=1)$. 
The relation between the  $M_k$'s and the $\tM_k$'s, and correspondingly between the $\mu_k$'s and the $T_k$'s, 
follows from the definitions, writing $(z^2\minus a^2)^{-k-\oh}$ as $(z^2-a_c^2 \plus \eps \sqL)^{-k-\oh}$ and expanding: 
\beq\label{jh11}
\mu_k = \sum_{l=k}^m c^{(k)}_{l-k} \;(l\plus \oh) T_l \;(\sqL)^{l-k} ,\quad k \geq 0, \quad 
c^{(k)}_{n} = (-1)^n \frac{\Gamma (n\plus k\plus \oh)}{\Gamma(k\plus \oh) n !},
\eeq
and inversely
\beq\label{jh11a}
(k\plus \oh)T_k = \sum_{l=k}^m (-1)^{l-k}c^{(k)}_{l-k} \; \mu_l \;(\sqL)^{l-k} ,
\eeq
where $c^{(k)}_{n}$ is the Taylor coefficients of $(1\plus x)^{-k-\oh}$. 

Eq.\ \rf{jh11} for $k=0$ determines $\sqL$ as 
a function of the $T_k$'s since $\mu_0 = 0$. Finally, the explicit linear relation between the coupling 
constants $T_k$ and the original coupling constants $t_{2k}$ can be expressed as follows  \cite{MMMM}:
\beq\label{jy1}
T_l= \eps^{l-m} \sum _{k\geq l} 
\frac{k t_{2k}\Gamma(k\plus \half)}{(k-l)! \,\Gamma(l\plus \frac 32)}, \quad l > 0,
\eeq
\beq\label{jy1a}
k t_{2k} = \sum _{l\geq k} (-1)^{k-l} \eps^{m-l}
\frac{T_l\Gamma(l\plus \frac 32)}{(l-k)! \, \Gamma(k\plus \half)}, \quad k >0.
\eeq
We note that differentiating \rf{jh3a} with respect to $T_0$ leads to a universal result only 
depending on the coupling constants via $\sqL$
\beq\label{jy2}
R_0(T) := -\frac{\prt}{\prt T_0} \Big(W_0(Z)-\oh \Vp(Z)\Big) = \frac{1}{2\sqrt{Z\plus \sqL}}
\eeq 
if one uses 
\beq\label{jy3}
\frac{\prt \mu_k}{\prt \sqL} = -(k\plus \oh)\;\mu_{k+1}, \qquad \frac{\prt T_0}{\prt \sqL} =  \, \mu_1, 
\eeq
which follow from  the definition of $\mu_k$ and from differentiating \rf{jh11} for $k=0$.
One recognizes \rf{jy2} as the scaling limit of \rf{jab4}.

The $w(z)$ from \rf{ja3W} admits a large $N$ expansion in powers of $1/N^2$, starting with $w_0(z)$. 
The so-called loop equation determines this expansion. It reads
\beq\label{jq5}
\oint_C \frac{\d \om}{2\pi \,i} \; \frac{\om V'(\om) \; w(\om)}{z\minus \om} = w^2(z) + \frac{1}{N^2}w(z,z),
\eeq
where $w(z_1,z_2)$ denotes the two-loop function. 

In the  scaling
limit defined by \rf{jh3} or \rf{jh10} this expansion takes the form 
\beq\label{jh6}
W(Z) = \sum_{h=0}^\infty (N \eps^{m+ \oh})^{-2h} \;  W_h(Z),
\eeq
where $W_h(Z)$ denotes the scaling limit  of the disk amplitude with $h$ handles. $W_h(Z)$ can be expressed in the following way \cite{ackm}, as a function of the 
$\mu_i$'s and $\sqL$:
\beq\label{jh7}
W_h(Z;\mu,\sqL) = \sum_{\a_i,n} C_h(\a_i,\a,n)\;
 \frac{\mu_{\a_1} \cdots \mu_{\a_k}}{\mu_1^\a }\; \frac{1}{(Z \plus  \sqL)^{n+ 3/2}}.
\eeq
In this formula the $1 <\a_i \leq m$, $\a \minus k = \chi(h)$ and $\sum_{i=1}^k (\a_i \minus 1) = 3g\minus (n\minus 1)$ 
so only a few 
$\mu_i$ contributes for low $h$ even if the multicritical index $m$ is very large. The constants $C_h(\a_i,\a)$
are rational numbers independent of the multicritical index $m$ and they are related to the 
intersection indices of a Riemann surface of genus $h$. Thus the only reference to the specific $m$'th multicritical point 
in \rf{jh3} is that $\mu_m \neq 0$ and $\mu_i =0$ for $i > m$. Still, in \rf{jh3} and \rf{jh6} we have an explicit reference to multicriticality in the genus expansion via
 the factor $N \eps^{m+\oh}$. It can be removed in the so-called {\it double scaling limit} if one introduces
 the coupling constant
 \beq\label{jh8}
 G^{-1} = N \eps^{m+ \oh}.
 \eeq

It is possible to rewrite \rf{jh7}  in terms of the $T_i$'s using \rf{jh11} in several ways:
\beq\label{jh12}
W_h(Z;T) = \sum_{n > 0} \frac{\hat{W}^{(h)}_{n} (T)}{(Z \plus \sqL(T))^{n+\oh}}=
 \sum_{n > 0} \frac{W^{(h)}_{n} (T)}{Z^{n+\oh}},
\eeq
where the Laurent series of $\sqZ \; W_h(Z;T)$ is convergent for large $Z$. 
The scaling limit of the loop equation \rf{jq5} can be written as 
\beq\label{jh15}
\oint_{C_1} \frac{d \Omega}{2\pi i} \; \frac{\Vp(\Omega)}{Z-\Omega} \; W(\Omega) = 
W^2(\Omega) + \frac{T_0^2}{8Z} + O\Big( \frac{G^2}{Z^2}\Big),
\eeq
where the contour $C_1$ encircles the cut of $ \Vp(Z) W(Z) $, i.e.\ the cut $[-\sqL,0]$, but not $Z$. Solving the loop equation with the boundary condition that $W(Z,T) 
= O(Z^{-3/2})$ determines $W(Z,T)$ as a function of $Z$ for given coupling constants $T$. 

A simple equation can be extracted from the $1/Z$ term in \rf{jh15}
\beq\label{jq6}
\oint_{C_1} \frac{d \Omega}{2\pi i} \; \Vp(\Omega)\, W(\Omega) =  \frac{T_0^2}{8}.
\eeq
One usually differentiate this equation with respect to $T_0$ to obtain the so-called {\it string equation}
\beq\label{jh23}
\oint_{C_1} \frac{d \Omega}{2\pi i} \; \Vp(\Omega) \; R(\Omega,T)  = 0,
\eeq
where $R(Z,T)$ denotes the generalization of $R_0(Z,T)$ defined by eq.\ \rf{jy2} to include higher genus contributions:
\beq\label{jh18}
R(T,Z) := \frac{1}{2\sqrt{Z} }-\frac{\prt}{\prt T_0} W(T,Z) = \sum_{n =0}^\infty \frac{R_n (T)}{Z^{n+\oh}}.
\eeq
This expansion is convergent for large $Z$ 
(as long as we restrict ourselves to a  finite order $h$ in \rf{jh12}). 

However it will also be 
convenient for us to consider other similar expansions 
\beq\label{jh18a}
R(T,Z) = \sum_{n =0}^\infty \frac{\hat{R}_{n} (T;a)}{(Z\plus a)^{n+\oh}} , 
\qquad  \hat{R}_n(T;a) =  \sum_{k=0}^n c^{(k)}_{n-k}\,(-a)^{n-k}R_k(T).
\eeq
When we consider the genus expansion of $R(T,Z)$ in powers of $G^2$ it is clear from 
\rf{jh12} that a natural choice is $a = \sqL$, but in the next section we will see that the KdV structure
of the loop equations also suggests another natural choice of $a$.
Deforming the integration $C_1$ to infinity,  using  that $\Vp(Z) R(Z;T)$ has a convergent 
Laurent expansion at $|Z| = \infty$, the string equation \rf{jh23} can be written
\beq\label{jh24}
\sum_{n=0}^m (n\plus \oh) T_n \;R_n (T) =0\qquad {\rm or} \qquad 
\sum_{n=1}^m \mu_n\; \hat{R}_n(T,a\! =\! \sqL)=0. 
\eeq

\subsection{Gelfand-Dikii and KdV hierarchy}
Let us define
\beq\label{jh21a}
D := \frac{d}{d T_0},\quad u(T):=R_1(T), \quad u'(T) := D u(T),~~u''(T) := D^2 u(T), ~~\cdots
\eeq
It follows from the loop equation \rf{jh15} that  (see e.g.\ \cite{book} for a review)
\beq\label{jh21}
\Big(G^2  D^3 +(u- Z)D + \oh (Du)\Big) R(Z;T) =0, \quad ,
\eeq
or in components 
\beq\label{jh22}
D R_{n+1} = \big(G^2D^3 + u D +\oh (Du)\big) R_n,
\eeq
or in a closed form 
\beq\label{jh22a}
R_n(u) = \Big( G^2 D^2 + \frac{u\plus D^{-1} u D}{2} \Big)^n \; \oh.
\eeq
The first few terms in the expansion are 
\beq\label{jh22b}
R_0=\frac 12,\quad R_1=\frac u4, \quad R_2 =\frac {G^2}{4} D^2 u+\frac 3{16}u^2,
\quad \ldots.
\eeq
$R(Z,T)$ thus only depends on the $T$'s via the function $u(T)$ and its derivatives after $T_0$ and the 
functions $R_n(u)$ are so-called Gelfand-Dikii (differential) polynomials. 

With this understanding 
we can write $R(Z,u)$ and this suggests that a natural expansion parameter $a$ in  eq.\ \rf{jh18a} is $a =-u$
and we can write
\beq\label{jh21d}
R(Z,u) = \sum_{n=0}^\infty \frac{\tR_n(u)}{(Z\minus u)^{n+\oh}},\quad \tR_n(u) =  \sum_{k=0}^n c^{(k)}_{n-k}\,u^{n-k}R_k(u),
\eeq
This expansion matches in a natural way with eq.\ \rf{jh21} which can be written as   
\beq\label{jh21aa} 
D \big( \sqrt{Z\minus u} \; R(Z,u) \big) =   \frac{G^2}{\sqrt{Z\minus u} }\; D^3  R(Z,u),
\eeq
and which provides us with an expansion of $R(Z,u)$ in powers of $G^2$.
Thus we can write $R(Z,u)$ as an expansion in $G^2$
\beq\label{jh21b}
R(Z,u) = \sum_{n=0}^\infty G^{2n} \tR^{(n)} (Z,u),
\eeq
and equation \rf{jh21aa} admits a recursive solution somewhat similar to the one shown in \rf{jh22}  
\beq\label{jh21c}
\tR^{(h+1)} (Z,u) =  \frac{1}{\sqrt{Z\minus u}}\;\Big(D^{-1} \frac{1}{\sqrt{Z\minus u}} \;D^3\Big)^h \tR^{(0)}(Z,u),
\quad \tR^{(0)}(Z,u) =    \frac{1}{2\sqrt{Z\minus u}}.
\eeq

We can solve \rf{jh21c}  iteratively and for further reference we give the two first terms
\bea
\tR^{(0)}&=&\frac{1}{2\sqrt{Z \minus u}} \label{jh21f1}\\
\tR^{(1)}&=&\frac{5 (u')^2}{16 \left(Z \minus u\right)^{7/2}}
+\frac{u''}{4 \left(Z\minus u\right)^{5/2}} \label{jh21f2}\\
\tR^{(2)} &=&
\frac{1155 (u')^4 } {256 \left(Z\minus u\right)^{13/2}}\plus \frac{231 (u')^2 u''}
 {32 \left(Z \minus u\right)^{11/2}}\plus
\frac{21  (u'')^2\plus 28  u' u'''} {16 \left(Z\minus u\right)^{9/2}}
\plus \frac{u''''}{4 \left(Z\minus u\right)^{7/2}}. \label{jh21f3} 
\eea
The dependence of $\tR^{(h)}(Z,u)$ on $Z$ will only  be via the functions $(Z\minus u)^{-n-\oh}$, where $n \in [h+1,3h]$
for $h >0$.
The expansion \rf{jh21b} is thus a rearrangement of the terms in \rf{jh21d} and we can conversely say that 
the numerator $\tR_n(u)$ in \rf{jh21d} will only contains powers $h$ of $G^2$ where $h\in \big[ [n/3]\plus 1,n\minus 1\big]$ 
for $n>1$.  
 
We should emphasize that \rf{jh21b} is not really a genus expansion from the point of view 
of the matrix model since $u(T)=R_1(T)$ itself has a genus expansion  determined 
by the string equation which, apart from being a differential equation in $T_0$ for $u(T)$, also can be 
solved iteratively in powers of $G^ 2$ (as we will discuss below). Let us write explicitly
\beq\label{jh21g}
u(T) = u_0(T) + G^2 u_1(T) +G^4 u_2(T) + \cdots.
\eeq
Then we have from \rf{jy2}, \rf{jh21f1} and \rf{jh21d} the genus zero result
\beq\label{jh27}
R^{(0)}(Z;T) = \frac{1}{2\sqrt{Z\plus  \sqL}}, \quad {\rm i.e.} \quad u_0(T) = - \sqL, \quad 
R^{(0)}_n (T) = \oh c^{(0)}_n (-u_0(T))^n,
\eeq
where $c^{(0)}_n$ is the coefficient in the power expansion of $1/\sqrt{1\plus z}$ and where $\sqL$ is a function 
of the $T_k$'s via eq.\ \rf{jh11}, namely $\mu_0 =0$.

Finally one can show (for a review see \cite{book}), again using the loop equation, that
\beq\label{jq11}
\frac{\prt}{\prt T_0} R_{n+1} = \frac{\prt u}{\prt T_n}.
\eeq
Together with \rf{jh22} this (potentially) infinite set of differential equations for $u(T)$ constitutes a so-called two-reduced
KdV-hierarchy of differential equations and it follows that (the singular part of) the partition function is the tau-function of 
this hierarchy \cite{KdV}.

\section{Generalized multicritical matrix models}\label{newstuff}

\begin{figure}
\begin{center}
\includegraphics[width=0.7\textwidth]{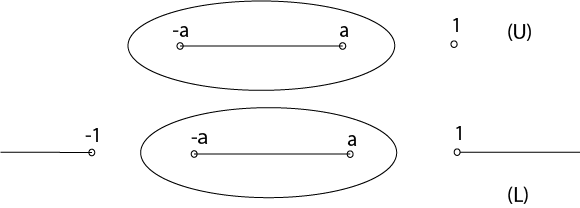}
\end{center}
\caption{For a polynomial potential, including the ordinary Kazakov potential,  
the cut in the complex $z$-plane is $[-a,a]$ and $a_c=1$ (by normalization) as illustrated on the 
upper figure (U). 
The contour can be anywhere outside the cut, and in particularly we can choose it to also include $\pm a_c$.
 For the generalized Kazakov the contour is pinched between $[a,a_c]$ and $[-a_c,-a]$ as shown on the 
 lower figure (L). These intervals shrink to points for $g \to g_*$.}
\label{cuts}
\end{figure}

In \cite{abm} we showed that eq.\ \rf{ja5} could be generalized to
\beq\label{jh28}
 g(a^2)= g_* - c (a_c^2 \minus a^2)^{s-\oh}, \quad g_* = c \,a_c^{2s-1}, \quad  c = \frac{1}{4(s-\oh) a_c^{2s-3}},
 \eeq 
 where $s \in ]1,\infty[$. By an allover scaling of the coupling constants one can always fix $a_c=1$ and we 
 will assume this in the following. While one could choose the potential $\tV(x)$ leading to \rf{ja5} as a polynomial 
 of order $2m$, \rf{jh28} with $s$ not a half-integer leads to a potential which is an infinite power series in $x$ where
 the coefficients $t_n$ fall of like $1/n^{s+1}$. Explicitly we found 
 \bea\label{jx2}
x \tV'_s(x)&=&
 {}_2F_1\Big(1,\frac32\minus s,\frac32, {x^2}\Big) x^2 \\
 &=& \frac{\pi \Gamma(s\minus \oh)}{2\Gamma(s)} x (1\minus x^2)^{s-1} +
 \frac{ {}_2F_1\left(1,\frac32\minus s,2\minus s, 1\minus x^2\right) x^2}{2(s\minus 1)}.
 \label{jxx2} 
\eea
The hypergeometric function appearing in \rf{jx2} shows that the power series of the potential has a 
radius of convergence  1 and cuts along the real axis for $|x| >1$, see Fig.\ \ref{cuts}. 
The representation \rf{jxx2} of $x \tV'(x)$ shows that the non-analytic structure of 
the potential in the neighborhoods of $x= \pm 1$ as well as along the cuts 
is entirely determined by the term $(1-x^2)^{s-1}$.

In the large $N$ limit we found the disk amplitude
 \beq\label{jh29}
 g w_0(z) - \oh \tV_s'(z) = -\oh \;\;
 {}_2F_1\!\Big(1, \frac32\minus s,\frac{3}{2},\frac{z^2\minus a^2}{1\minus a^2}\Big)
(1\minus a^2)^{s-\frac{3}{2}} \sqrt{z^2\minus a^2}
\eeq 
where the relation between $a$ and $g$ is given by \rf{jh28},  i.e.
\beq\label{ja10}
a^2 = 1 - \left(1\minus  g/g_*\right)^{1/(s-1/2)}, 
\quad g_* = \frac{1}{4 (s \minus 1/2)}.
\eeq

It is seen that the analytic structure of $ g w_0(z) \minus \oh \tV_s(z)$ is very similar to the one given 
by \rf{jh1a}, except that the function $M(z)$ is not a polynomial but an infinite power series.
It allows us directly to take the scaling limit \rf{jh3b} and we obtain 
\beq\label{ja10a}
g w_0(z)-\oh \tV'_s(z)
=-\frac 12 \epsilon^{s-1} \left[(\sqL)^{s-\frac{3}{2}}  {}_2F_1 \Big(1,\frac{3}{2} \minus s, \frac 32; 
\frac{Z\plus \sqL}{ \sqrt{\Lambda}}\Big)\right]\;\sqrt{Z\plus  \sqL}.
\eeq
The hypergeometric function has a power expansion in $Z \plus \sqL$ and it is seen that 
this is precisely the expansion in \rf{jh3a} with $\mu_k$ given in \rf{jh30} below.
In order to connect it to $W_0(Z)$ and $\Vp(Z)$ in \rf{jh3a} we perform a Kummer
transformation of the hypergeometric function and obtain
\bea\label{jh10b}
 \lefteqn{g w_0(z)-\oh \tV'_s(z) = } \nonumber \\
 &&
 -\eps^{s-1} \left(
(\sqrt{\Lambda})^{s-\frac 12} \; \frac{ \sqrt{ Z \plus \sqrt{\Lambda}}}{4(s-\oh)Z}\;
{}_2F_1 \Big(1,s, \frac 12 \plus s; -\frac{\sqrt{\Lambda}}{ Z}\Big) +
 \frac{\sqrt{\pi} \Gamma(s\minus \frac 12)}{4 \Gamma(s)} \;
(-Z)^{s-\frac{3}{2}}\sqZ\right).  \nonumber \\
&&
\eea
From this we conclude that 
\beq\label{jh10c}
W_0(Z) = \frac{(\sqrt{\Lambda})^{s-\frac 12}}{2} \;\left( -\frac{ \sqrt{ Z \plus  \sqrt{\Lambda}}}{Z}\;
{}_2F_1 \Big(1,s, \frac 12 \plus s; - \frac{ \sqrt{\Lambda}}{ Z}\Big) +\frac{1}{\sqZ} \right)
\eeq
since this is the part which has an expansion in $Z^{-n-\oh}$, $n >0$. The rest is now 
identified with $\Vp(Z)$:
\beq\label{jh10d}
\Vp(Z) = \frac{(\sqL)^{s-\oh}}{\sqZ} -\frac{\sqrt{\pi} \Gamma(s\plus \frac 12)}{ \Gamma(s)} \;
\frac{(-Z)^{s-\oh}}{\sqZ},
\eeq
i.e.\ we can write (making a somewhat arbitrary choice of how to divide $\Vp(Z)$ into $\tT_{s-\oh}$ and a $Z$ 
part for $s\minus \oh$ not an integer, the ambiguity caused by the cut along the positive real axis)
\beq\label{jh10e}
\oh T_0 = \oh \tT_0=  (\sqL)^{s-\oh},\qquad s\,\tT_{s-\oh} =  -\frac{\sqrt{\pi} \Gamma(s\plus \frac 12)}{ \Gamma(s)}.
\eeq

It is seen by comparing with \rf{jq1} that the potential in \rf{jh10d}
does {\it not} have an expansion in powers $Z^{k+\oh}$. Let us try to 
understand why this is so, and how one can reach ordinary $T_k$'s as
in the expansion \rf{jq1}.

Like the ordinary Kazakov potential, the so-called generalized Kazakov potential \rf{jx2} only contains
one adjustable coupling constant $g$, and the position of the cut of $w_0(z)$ is determined by \rf{ja10}.
The difference is that the derivative of the potential has  cuts along the real axis for $ |z| >1$.
This implies two things. Firstly we can still define the moments $M_k$ like in \rf{jh1}. We just have to 
choose the contour such that it encloses the cut  $[-a,a]$ and avoids the cuts $|z| > 1$ along the real axis.
Once we make a deformation $t_n \to t_n \plus \delta t_n$ which leads to a scaling \rf{jh3}, the positions of 
the two cuts will change, but as long as they stay separated (which we will assume and 
examples of which we will
provide below) we will have no problem with the choice of contour. 
This implies that almost all statements
made when discussing the $m^{{\rm th}}$ multicritical matrix model in Sec.\ \rf{oldstuff}  which relate to 
the $\mu_i$ will still be valued. In particular \rf{jh7} will remain valid.
The ``only'' difference is that we now have infinitely many moments different from zero when we 
approach the critical point and that the higher moments become singular in the scaling limit. 

This is 
true even for the plain generalized Kazakov model where we only have one coupling constant to adjust. 
In that case we have explicitly
\beq\label{jh30}
M_k = d^{(s)}_k (1\minus a^2)^{s-\oh -k} = \mu_k \eps^{s-\oh -k}, 
\quad \mu_k = d^{(s)}_k (\sqL)^{s-\oh-k},\quad k >0,
\eeq
where 
\beq\label{jh30a}
d^{(s)}_k = (-1)^{k-1}\;\frac{(s\minus \oh)\cdots (s\minus (k\minus \oh))}{\oh \cdots (k\minus \oh)} = 
- \frac{\Gamma(k\plus \oh \minus s) \Gamma(\oh)}{\Gamma(\oh \minus s)\Gamma(k \plus \oh)}.
\eeq
{\it Despite the increasing singular behavior of $M_k$ with increasing $k$, this behavior does not 
appear in  \rf{jh7}.} However, for $s\minus \oh$ non-integer the $T_k$'s defined by \rf{jh11a} are
\beq\label{jh30b}
\oh T_0 = (\sqL)^{s-\oh}, \quad T_k = 0, ~~0 < k < s-\oh, \quad T_k = \infty ~~k > s-\oh.
\eeq  
This singular behavior of the $T_k$'s is also seen in \rf{jy1} where $T_l$ is divergent for $l > s-\oh$ provided
$t_{2k} \propto 1/k^{s+1}$ for large $k$. This asymptotic behavior of the $t_{2k}$ results in a singular behavior 
of $xV'(x) \sim (1\minus x^2)^{s-1}$ and from the representation \rf{jh9}, using the last expression in eq.\ \rf{jh1}
for $M_k(t;a)$,  it is indeed seen that such a 
behavior implies that $\tM_k$ becomes infinite for $k > s\minus \oh$. 

This problem reflects that for a potential 
with the singularity structure like that of the generalized Kazakov  potential \rf{jx2} we cannot 
without qualifications use the definitions  \rf{jh9} and \rf{jh10} for $\tM_k$ and $T_k$. The reason is that 
the contour integral in \rf{jh9} had to encircle the cut $[-1,1]$, but the Kazakov potential has a cut on the 
positive real axis, starting from $z=1$. Stated differently,  in order for \rf{jq1} to make sense  there has to 
be a region in the complex plane such that $(W_0(Z) - \oh \Vp(Z))/\sqrt{Z}$ has a Laurent expansion. For a
standard polynomial potential $V(z)$ the neighborhood of infinity always allows such an expansion. In the case 
of the generalized Kazakov potential there is no such region as is clear from the expressions \rf{jh10b}-\rf{jh10d}.
We can solve this problem by considering  deformations $t_n \to t_n \plus \delta t_n$
which still have $a_c \!=\!1$ but where the cut of $V'(z)$ approach $z\!=\!1$ from above when $\delta t_n \to 0$.
Then we can define the $\tM_k$ and $T_k$
and everything stated with respect to the KdV-times $T_k$ will remain valid. 
To be explicit we will consider a class of deformations 
which are general enough to allow for an arbitrary finite number of {\it independent}\/ KdV-times $T_k$ different from zero.

\begin{figure}
\begin{center}
\includegraphics[width=0.7\textwidth]{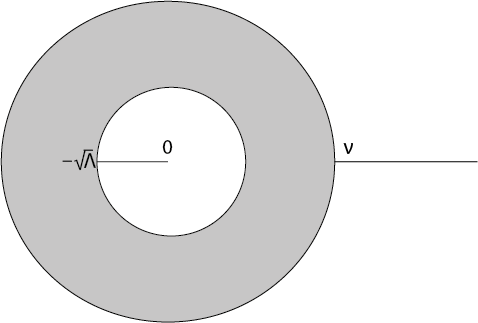}
\end{center}
\caption{The annulus where $(W_0(Z)\minus  \oh \Vp(Z,\nu))/\sqrt{Z} $ has a Laurent expansion. The right cut 
is from  $\Vp(Z,\nu)/\sqZ$ given by \rf{jh34} and starts at $\nu$, while the cut from 0 to $-\sqL$ is from 
$\sqrt{Z+\sqL}/\sqZ$.}
\label{annulus}
\end{figure}

Let us start with the $t_n$'s corresponding to the generalized Kazakov potential \rf{jx2}. In this case we have one coupling 
constant which can be adjusted, $g$. We now introduce another adjustable parameter $\eta$ by making the deformation  
\beq\label{jh31}
t_n \to t_n \eta^{2n} = t_n + \delta t_n , \quad \eta^2 = 1 - \eps \nu < 1,\quad {\rm i.e.} \quad V_K(z) \to V_K (\eta z),
\eeq
which has a cut starting at $z = 1/\eta  \approx 1\plus  \oh \eps \nu$.  Following \rf{jh30} we now have moments 
\beq\label{jh32} 
M_k = d^{(s)}_k\eta^{2k} (1\minus \eta^2 a^2)^{s-\oh -k},  \quad 
\mu_k = d^{(s)}_k (\sqL \plus\nu)^{s-\oh -k}, \quad k>0,
\eeq
and we obtain from \rf{jh9} in analogy with \rf{jh30} 
\beq\label{jh33}
\tM_k  = d^{(s)}_k (1\minus \eta^2)^{s-\oh -k},\quad {\rm i.e.}\quad
 (k\plus \oh)T_k = d^{(s)}_k \nu^{s-\oh -k}, \quad k>0.
\eeq
For $k=0$ we have 
\beq\label{jh33a}
M_0 = \frac{g_*}{g}\big(1-\eps^{s-\oh}(\sqL\plus \nu)^{s-\oh}\big)
\qquad \tM_0 =  \frac{g_*}{g}\big(1- \eps^{s-\oh}\nu^{s-\oh}\big),
\eeq
which tell us (since $M_0 =1$) that 
\beq\label{jh33b}
\frac{g_*}{g}= 1 +\eps^{s-\oh}(\sqL\plus \nu)^{s-\oh},\qquad \oh T_0 = (\sqL\plus \nu)^{s-\oh}- \nu^{s-\oh}.
\eeq
The relation between $\mu_k$ and $T_k$ is given by \rf{jh11}, but while the summation in \rf{jh11} is finite 
for a polynomial potential, it will not terminate in the present case.  Again to be explicit, the corresponding 
continuum potential is 
\beq\label{jh34}
\Vp(Z,\nu) = \sum_{n=0}^\infty (n\plus \oh) T_n Z^{n-\oh} = \frac{(\sqL\plus \nu)^{s-\oh}}{\sqZ} - 
\frac{\nu^{s-\oh}}{\sqZ} \;{}_2F_1\Big(1, \frac12\minus s,\frac{1}{2},\frac{Z}{\nu}\Big)
\eeq
and $\Vp(Z,\nu)\,\sqrt{Z}$ has a convergent power series for $|Z| < \nu$. Furthermore 
\beq\label{jh35}
W_0(Z)- \oh \Vp(Z,\nu) = -(s\minus \oh)\; \; {}_2F_1\Big(1, \frac32\minus s,\frac{3}{2},\frac{Z\plus \sqL}{\nu\plus \sqL}\Big)
\big(\nu\plus \sqL\big)^{s-\frac{3}{2}} \sqrt{Z\plus \sqL},
\eeq
and for  $\sqL < |Z| < \nu$ there is a convergent Laurent expansion of   $(W_0(Z)\minus  \oh \Vp(Z,\nu))/\sqrt{Z} $
(see Fig.\ \ref{annulus}).
By a suitable Kummer transformation one can show that $W_0(Z)/\sqrt{Z}$ for sufficient large $Z$ 
has a convergent expansion in $1/Z$, starting with the power $1/Z^2$.

It is now clear that we can obtain an arbitrary (finite) number of {\it independent $\mu_k$ and $T_k$} by choosing 
a potential 
\beq\label{jh34a}
V(z) = \sum_{a=1}^L c_a V_K(\eta_a z), \quad (k\plus \oh)\;T_k =  \sum_{a=1}^L c_a d_k^{(s)}\,\nu_a^{s-\oh -k},
\quad \sum_{a=1}^L c_a =1,
\eeq
and in this way we obtain a KdV hierarchy expressed in terms of the standard KdV times $T_k$. 
However, in general an infinite number of $T_k$ will be different from zero (but not independent 
of the $T_k$ with lower values of $k$).
This implies that contrary to the situation for ordinary matrix models approaching a multicritical 
point of order $m$ the string equation \rf{jh24} is now an infinite order differential equation in $T_0$ for 
the determination of $u$ and in this sense only practical useful when viewed as a genus expansion,
in which case only a finite number of derivatives appear to each order. While the representation of
perturbations around the generalized Kazakov potential is somewhat singular in terms of the standard
KdV times $T_k$, it should be noted, and this will be used below, that the time $T_0$ is alway well defined 
and non-singular and the same is true for the function $u(T)$ and  the 
representation in terms of Gelfand-Dikii  polynomials $R_n(u)$.

\section{The string equation}\label{GD}    

As seen above, contrary to the $\mu_k$, the standard KdV times $T_k$ 
are maybe not the most natural variables  when we consider deformations away from the  
generalized Kazakov potential. This is not  surprising. Recall that for the $\mth$ multicritical Kazakov
potential the anchor point was really $T_m \neq 0$ (and the rest of the $T_k$ equal zero, $k>0$).  
For the generalized Kazakov we had $T_k =0$ for $0 < k < s-\oh$ and $T_k = \infty$ for $k> s -\oh$). 
What naturally replaced $T_m \neq 0$ was $\tT_{s-\oh} \neq 0$ as seen in \rf{jh10d}.

In the $\mth$ Kazakov model deformations away from the pure model resulted in $T_k$ different 
from zero for $k=1,\ldots,m-1$ and these deformations could be viewed in a Wilsonian renormalization language as 
relevant deformations away from the multicritical point.  Like in a renormalization group flow the ``bare coupling 
constants'', the $\tM_k$, have to be turned to zero (like $\eps^{m-k}$). If $\tM_k$ is not tuned to zero
the system will ``flow'' away from the neighborhood of the $\mth$ multicritical point and to the $k^{{\rm th}}$ multicritical point
when the ``cutoff'' $\eps$ is taken to zero, the lowest $k >0$ where $\tM_k \neq 0$ dominating. 
From this perspective,
given a $\mth$ multicritical point, the $T_k$, $k>m$ correspond to the (infinite) set of irrelevant deformations which 
naturally scale to zero when we approach the critical point (formerly $T_k = \eps^{k-m} \tM_k$, $\tM_k$ finite). 
Clearly the $T_k$ we have introduced above by deforming the potential away from the generalized Kazakov 
critical point corresponding to an index $s \neq m \plus \oh$ are slightly different in nature since we have 
infinite many of them, so the concept of relevant and irrelevant deformations does not apply in the same way.
Maybe this is not surprising since the deformations of the $\mth$ multicritical point and the 
$T_k$  have an interpretation in  Liouville quantum gravity where they can be related to deformations 
corresponding to the $m$ (gravity dressed) primary operators  in the 
$(2,2m\minus 1)$ conformal field theory coupled to quantum gravity. 
However, for $s$ not a half-integer and in general 
an irrational number, the corresponding\footnote{In \cite{abm} we showed how one formally can associate a central 
charge $c(s)$ to a given $s$. For some range of $s$ there is good evidence for this formal 
relation, However, for general $s$ the relation is just based to calculating the ``string'' susceptibility exponent  
$\gamma(s)$ and using the KPZ relation between $\gamma$ and the central charge $c$. See \cite{abm} for a 
discussion of these issues.}
conformal field theory will in general be irrational and there is no 
obvious finite  set of primary operators which can be considered as the relevant deformations. 
Thus a more natural starting point might be to consider a given 
generalized Kazakov potential $V_s(x)$ corresponding to an $s\neq m\plus \oh$ and then 
consider ``relevant'' deformations  
\beq\label{jh55}
V(x) = \sum_{0\leq k < s-\oh}  \tit_{s-k} V_{s-k} (x), \quad \tit_{s-k} = O(\eps^k ).
\eeq
Such a potential will result in a continuum potential where 
\beq\label{jh56}
\Vp(Z) = \sum_{0\leq k < s-\oh} (s-k) \tT_{s-\oh-k} (-Z)^{s-\oh -k}/\sqZ
\eeq
and as mentioned it would be the ``naive'' generalization of the relevant deformations around
the Kazakov potential. 

One can view it has a formal decomposition of any 
polynomial $V(x)$, changing coupling constants from $t_k$ to coupling constants $\tit_s$ in 
the same way as we could formally change from $t_k$ to $\eps^{m-k}T_k$ via formula \rf{jy1}.%
\footnote{The explicit formula corresponding to \rf{jy1a} is 
\beq\label{jyy1}
k t_{2k}= \sum_{n=1} ^\infty (-1)^{n-1} \frac {\eps^{-n-\alpha}}{{G}}
\frac{\wt T_n \,\Gamma(n+\alpha+1)\,\Gamma(k-n-\alpha+\half)}{\Gamma(n+\alpha-\half)\,\Gamma(k+\half)\,\Gamma(\frac 32- n -\alpha)}.
\eeq
It hold for arbitrary $\alpha = s-[s] \in [0,1[$ and reproduces \rf{jy1a} 
 for $\alpha=1/2$.
}
A priori it is not clear why a decomposition like \rf{jh55} and correspondingly \rf{jh56} should be chosen 
for an $s$ where $s-\oh$ is not an integer. It appears just to be an imitation of the situation 
for $s-\oh = m$, where there are good reasons for such an expansion, as explained above. Why not 
integrate over some range of $s$ since $s$ is a continuous variable rather than the discrete $m$? 
We can obtain some evidence in favor of the choice \rf{jh56} by looking at the deformed potential 
\rf{jh34}. By a Kummer transformation the large $Z$ behavior of the potential is 
\beq\label{jh34aa}
 \Vp(Z,\nu) = -\frac{\sqrt{\pi} \Gamma(s \plus \oh)}{\Gamma(s)} \; \frac{(-Z)^{s-\oh}}{\sqZ} \Big(1\minus\frac{\nu}{Z}\Big)^{s-1} 
 + 0(Z^{-\oh}).
 \eeq
 Thus  the potential $\Vp(Z,\nu)$  indeed has a expansion like \rf{jh56} in terms of potentials \rf{jh10d} with $s, s\minus 1, \ldots$.

Let us now study in more detail the genus expansion associated with the simplest potential 
\rf{jh10d}, the  generalized Kazakov potential $V_s(x)$. The results can easily be extended
to the more general potential \rf{jh56}. As mentioned above, 
formulated perturbatively and in terms of the moments $\mu_k$,
everything is as for ordinary matrix models, except that we have infinitely many $\mu_k$ different from 
zero and explicitly given by \rf{jh30}. We thus have a well defined genus expansion. In fact, since 
we know all $\mu_k$ for the generalized Kazakov potential (see \rf{jh30}) we can just insert them in the 
standard expression for the genus expansion, e.g.\ following \cite{ackm}. 

There is however another way 
to obtain the genus expansion  i.e.\ for the disk amplitude with one puncture with respect to $T_0$, $R(Z,u(T))$, 
namely first to solve the string equation for $u(T)$ and then to use \rf{jh18} and \rf{jh22a}. For the 
$m$'th multicritical model we have 
\beq\label{jh57}
\oh T_0 = (\sqL)^m,\qquad (m\plus \oh)T_m = \frac{(-1)^{m-1} \Gamma(\oh) \Gamma(m\plus  1)}{\Gamma(m+\oh)},
\eeq
and the contour in the string equation \rf{jh23} can be deformed to infinity to give \rf{jh24}, which 
in the present case reads
\beq\label{jh58}
(m\plus \oh)T_m R_m(u) + \oh T_0 R_0 =0,\qquad \oh T_0 = (\sqL)^{m}.
\eeq  
In principle we can solve this finite order non-linear  differential equation for $u$ 
and the solution has the  form
\beq\label{jh60} 
 u(T_0,G) = -\sqL + \sum_{h=1}^\infty c_h \Big(\frac{G^2}{\Lambda^m}\Big)^h
=-\Big(\frac{T_0}{2}\Big)^{\frac{1}{m}} +\sum_{h=1}^\infty c_h \Big(\frac{4G^2}{T_0^2}\Big)^h . 
\eeq 
 If we  only are interested in the perturbative genus expansion in powers of $G^2$ equation \rf{jh58} allows
for a simple iterative algebraic determination of the coefficients $c_h$. 

For the generalized Kazakov potential we have $\oh T_0 = (\sqL)^{s-\oh}$ but  no longer a finite order 
differential equation which determines $u(T_0,G)$. However we  still have a simple algebraic 
procedure to determine the genus expansion of $u(T_0,G)$. In this expansion 
the equivalent of  \rf{jh60} reads:
\beq\label{jh60a} 
u(T_0,G) = -\sqL + \sum_{h=1}^\infty c_h \Big(\frac{G^2}{\Lambda^{s-\oh}}\Big)^h =
 -\Big(\frac{T_0}{2}\Big)^{\frac{1}{s-\oh}} + \sum_{h=1}^\infty c_h \Big(\frac{4G^2}{T_0^2}\Big)^h.
\eeq 

The string equation \rf{jh23} is still valid for the generalized Kazakov potential. We can think of the string 
equation as derived from the $\nu$ deformed potential \rf{jh34} in the limit $\nu \to 0$. 
If we start with a situation where $\nu > \sqL$ we have a standard Laurent expansion, as explained above.
Since we know the $T_k$ (eq.\ \rf{jh33}), we can immediately write 
\beq\label{jm1}
\sum_{k=1}^\infty d^{(s)}_k \nu^{s-k-\oh} R_k(u) + \oh T_0 R_0 = 0, \qquad \oh T_0 =  (\sqL +\nu)^{s-\oh} - \nu^{s-\oh}.
\eeq
This infinite order differential equation looks singular for $\nu \to 0$. However, using 
\beq\label{jm2}
R_k(u) = \sum_{l=0}^k c^{(l)}_{k-l} (-u)^{k-l} \tR_l(u), \qquad c^{(l)}_{k-l} d^{(s)}_k = c^{(l-s)}_{k-l} d^{(s)}_l,
\eeq 
where the first equation is the inverse of the equation in \rf{jh21d} and where $c^{(k)}_n$ and $ d^{(s)}_k$ are 
defined by \rf{jh11} and \rf{jh30}, we obtain (for $\nu > |u|$)
\beq\label{jh64}
\sum_{k=0}^\infty d^{(s)}_k (\nu-u)^{s-\oh-k} \tR_k(u) + \oh  (\sqL +\nu)^{s-\oh} = 0.
\eeq 
This equation is regular in the limit $\nu \to 0$ and the final string equation is thus
\beq\label{jh64a}
\sum_{k=0}^\infty d^{(s)}_k (-u)^{s-\oh-k} \tR_k(u) + \oh T_0 \tR_0  = 0, \qquad \oh T_0 = (\sqL )^{s-\oh},
\eeq
and for genus 0 it gives precisely $-(-u)^{s-\oh} +\oh T_0=0$, i.e. $u= -\sqL$.

The string equation could have been derived simpler starting from the 
expansion \rf{jh21d}. Again taking $\nu > 0$ the contour $C_1$ exists and will enclose the cut $[u,0]$ and one obtains
\beq\label{jh64b}
\lim_{\nu \to 0}\oint_{C_1} \frac{d \Omega}{2\pi i} \; \Vp(\Omega,\nu) \; \frac{\tR_k(u)}{(\Omega\minus u)^{k+\oh}} =
d_k^{(s)} (-u)^{s-\oh-k} \tR_k(u),\quad k>0,
\eeq
while the integral for $k=0$ is $\oh\big((\sqL)^{s-\oh}\minus  (-u)^{s-\oh}\big)$. The result is obtained 
either by contracting the integral to the cut or deforming it to infinity (details are provided in the Appendix).
The string equation follows immediately and in this derivation  we do not have to 
assume $\nu > |u|$ as a starting point. 

As mentioned, in general the string equation \rf{jh64a} will be an infinite order differential equation.
Since $d^{(s)}_k\!=\!0$ for $s\minus \oh = m$ and $k > m$  the infinite series in \rf{jh64a} will in this case 
terminate we recover the string equation \rf{jh58} for the $m$'th multicritical matrix model. However,
viewed as a genus expansion of $u$ in powers of $G^2$ eq.\ \rf{jh64a} is not much different from the ``ordinary"
string equation.  As mentioned already below eq.\ \rf{jh21f3}, if we want to solve \rf{jh64a} up to order $h$, using 
the expansion \rf{jh21g}, it will involve only the first $3h$ terms in \rf{jh64a} and to obtain these we
only have to solve iteratively eq.\ \rf{jh21c} up to order $h$. Using \rf{jh21f1}-\rf{jh21f3} the string equation 
\rf{jh64a} to order $G^4$ reads:
\bea
 \lefteqn{\frac{1}{4} T_0= 
 (-u)^{s-1/2}\oh +G^2(-u)^{s- 5/2}\;\frac{2^2(s\minus \frac12)(s\minus \frac32)}{1\cdot 3} 
 \left[ \frac{ u^{(2)}}{4}   + 
\frac{2(s\minus \frac52)}{5\cdot u} \frac{(u')^2}{16} \right]} \nonumber \\
&&+ G^4 (-u)^{s-\frac72}\;    \frac{2^3 (s \minus \frac12)(s\minus \frac32)(s \minus \frac52)}{1\cdot 3\cdot 5}
\left[ \frac{u^{(4)}}{4} +\frac{2(s-\frac72)}{7\cdot u} \Big( \frac{28 u'u^{(3)}\plus 21 (u^{(2)})^2 }{16}\Big)\right.\non
&&+\frac{2^2 (s\minus \frac72)(s\minus \frac92)}{7 \cdot 9 \cdot u^2} \Big( \frac{231}{32} (u')^2u^{(2)} \Big)
\left.+\frac{2^3(s\minus \frac72)(s\minus \frac92)(s\minus \frac{11}{2})}{7 \cdot 9 \cdot 11 \cdot  u^3} 
\frac{1155}{256} (u')^4 \right]
\label{45a}
\eea  
which allows us to determine the coefficients $c_h$ in eq.\ \rf{jh60a} for the expansion  of $u$ in powers of $G^2$.

\section{Discussion}\label{discussion}

We have studied perturbations around and the genus expansion of the generalized Kazakov model 
we introduced in \cite{abm}. The model serves as a prototype for matrix models where the 
coefficients in potential  $V(x) = \sum_n t_n x^n$ behave as $t_n \sim 1/n^{s+1}$ for $n \to \infty$.
When the pertubation away from the Kazakov potential $V_s(x)$ was formulated in terms of the 
so-called moments $M_k$ and $\mu_k$, everything was a simple generalization of what happens for polynomial
potentials, except for the the fact that we have infinitely many moments $M_k$ and $\mu_k$ different from 
zero, contrary to the polynomial case where perturbations around the $\mth$ multicritial point only involve
moments $\mu_1,\ldots,\mu_m$ in the scaling limit. Despite the fact that the moments $M_k$ become
increasingly singular in the scaling limit for $k > s\minus \oh$, the standard formulas for polynomial potentials
valid for any polynomial interactions as well as to any order in the genus expansion are still valid. The 
 increasingly singular behavior of the moments $M_k$ cancel in the expressions for multiloop 
correlations leaving us with finite expressions for the multiloop correlators when expressed in terms of 
the scaled moments $\mu_k$. These expressions are identical to the expression 
for a standard multicritical matrix model with a sufficiently high $m$ 
(recall that for a given multiloop function and expanded to a given order,
only a finite number of  $\mu_k$ will appear). 

However, the situation was less clear when it came to the relation to the KdV-structure of ordinary, polynomial
one-matrix models. The problem is that the continuum potential in the case of the generalized Kazakov model
does not have an asymptotic expansion $\sum_n T_n Z^{n+\oh}$ which allows us to define the continuum
coupling constants $T_k$ which serve as KdV-times in the KdV hierarchy equations. We have shown that 
by a suitable deformation away from the pure generalized Kazakov model one can indeed define such times.
One important difference to the standard KdV situation is however that at least for the class of deformations 
we consider, one always has an infinite set of $T_k$ which are not independent. We showed that one 
can make an arbitrary finite set of these $T_k$ independent, but a basic property of the KdV hierarchy, namely
that it forms a closed system of differential equations for any finite set of $T_k$ is not true for our system. 
We will have a system of differential equations of infinite order where one can choose arbitrarily a finite 
set of {\it independent} $T_k$, but this is somewhat different from being able to choose a finite set of $T_k$.
Assuming that there still is an integrable structure behind these equations, this suggests one should be able to 
choose a different set of times $\tT_k$  by extracting the singularity in the series expansion
 of the continuum potential $\sum_n \tT_n Z^{n+\alpha}$  with $\alpha=s\minus[s]$,
where the standard KdV properties are satisfied. We suggested such 
a set, but it is not yet clear if they can serve as genuine KdV times.

Finally we addressed the question of the genus expansion of $R(Z,u)$ for the generalized Kazakov model.
As mentioned above the genus expansion is unproblematic when formulated in terms of the 
moments $\mu_k$. However, the genus formulation for $R(Z,u)$ is
usually nicely formulated via the string equation which (in principle) can be solved for 
$u(T)$ to all orders in the genus expansion, after which one can find the genus expansion of $R(Z,u)$.
Since the string equation in terms of the $T_k$ is not available for us for the generalized 
Kazakov model (no $T_k$, $k >0$), we investigated in detail the string equation for the 
generalized Kazakov model and showed how the string equation still led to a perturbative 
determination of the function $u(T_0)$ via an equation which reduced to the standard finite order
differential equation for $s\minus \oh =m$. We also showed that the string equation 
is nicely formulated via the times $\tT_k$ when $s$ is not half-integer.

\subsection*{Acknowledgments}

We are grateful to Timothy Budd for useful discussions at early stages of this work.
The authors acknowledge  the support by  the ERC Advanced
grant 291092, ``Exploring the Quantum Universe'' (EQU)
as well as the Dainish Research Council grant  ``Quantum Geometry''. 
L.~C.\ and Y.~M.\  thank the Theoretical Particle Physics and Cosmology group 
at the Niels Bohr Institute for the hospitality.

\section*{Appendix}

Let $\nu >0$. Thus the deformed potential $\Vp(Z,\nu)$  
is analytic in the region $Z \in ]0,\nu[$ and the contour integral 
\beq\label{app1}
I_k(u,\nu) := \oint_{C_1} \frac{d \Omega}{2\pi i} \;   \frac{ \Vp(\Omega,\nu)}{(\Omega\minus u)^{k+\oh}} 
= \frac{1}{\Gamma(k\plus \oh)} \frac{d^k I_0(u,\nu)}{du^k} 
\eeq
around $C_1$  is well defined and can be calculated by contracting the contour to 
the cut $[u,0]$. Once this is done there is no problem taking $\nu \to 0$ and we end up with 
the integral 
\beq\label{app2}
\int^0_u \frac{dx}{\pi} \; \left( \frac{(\sqL)^{s-\oh}}{\sqrt{-x}} - \frac{\sqrt{\pi} \Gamma(s+\oh)}{\Gamma(s)} \; 
\frac{(-x)^{s-\oh}}{\sqrt{-x}}\right) \; \frac{1}{\sqrt{-u -x}} = (\sqL)^{s-\oh} - (-u)^{s-\oh}.
\eeq
We thus have  
\beq\label{app3}
I_0(u,0) =  (\sqL)^{s-\oh} - (-u)^{s-\oh},\qquad I_k(u,0) = d^{(s)}_k (-u)^{s-\oh -k}.
\eeq

If we instead deform the contour to infinity and along the cut of $\Vp(Z,\nu)$ on the positive real axis starting at $\nu$ 
we can express the string equation in terms of the Mellin transform of $R(Z,u)$. Again, after the deformation
we can take $\nu \to 0$ and we will assume this is done in the following. For $k > s\minus \oh$ the contribution 
to $I_k(u,0)$ from the contour at infinity will vanish and we have left with the contribution from the cut:
\beq\label{app4}
I_k(u,0) = \cos \pi s \int_0^\infty \frac{dx}{\pi} \frac{\sqrt{\pi} \Gamma(s+\oh)}{\Gamma(s)} \frac{x^{s-1}}{(x\minus u)^{k+\oh}} . 
\eeq
For $k < s\minus  \oh$ there is a contribution from the contour at infinity and this contribution 
exactly cancels  the contribution from the integral in \rf{app4} for large    $x$ since $I_k(u,0)$ of course
is independent of the choice of contour. We can thus view \rf{app4} as valid for all $s$ provided 
we  subtract from the integrand a sufficient number of terms of the Taylor expansion around
infinity to ensure convergence of the integral at infinity. This is exactly what is done for Mellin transformations
defined as follows
 \beq\label{app4a}
\tilde{f}(s):= \cM[f](s) = \int_0^\infty dx \; x^{s-1} f(x),\quad  
\cM^{(-1)}[\tilde{f}](x) = \int_{\gamma -i\infty}^{\gamma+i\infty} ds \; x^{-s} \tilde{f} (s),
\eeq 
if the integral is divergent at infinity. We can thus write
\beq\label{app5}
\cM[(x-u)^{-k-\oh}] (s) = (-u)^{s-\oh -k} d^{(s)}_k \;\frac{\Gamma(s) \Gamma(\oh \minus s)}{\Gamma(\oh)},
\eeq
and applying to $R(Z,u)$ we have
\beq\label{app5a}
\cM[R(\cdot,u)](s) := \int_0^\infty dx x^{s-1} R(x,u) = 
\frac{\Gamma(s) \Gamma(\oh \minus s)}{\Gamma(\oh)}\sum_{k=0}^\infty d^{(s)}_k (-u)^{s-\oh-k} \tR(u).
\eeq
We can thus write the string equation as 
\beq\label{app6}
 \cM[R(\cdot,u)](s) = -T_0  \frac{\Gamma(s)\Gamma(\oh\minus s)}{2\sqrt{\pi} },
 \eeq
 where the term on the rhs comes from the $\oh T_0/\sqZ$ part of the potential $\Vp(Z)$ in
 the string equation.  This equation looks intriguing simple and one could be tempted
 to apply the inverse Mellin transform to find $R(x,u)$ since the inverse Mellin
 transform of the rhs is just%
 \footnote{This is true for $0< s <\oh$ where the Mellin transform 
 of $R_0(x,-\sqL)$ exists. For larger $s$ the inverse Mellin transform will be 
 \beq\label{js1}
 R_0(x,-\sqL) -\sum_{k=0}^{k_0} c^{(0)}_k \frac{(\sqL)^k}{x^{k+\oh}},\qquad k_0 = [s-\oh],
 \eeq
 which has the rhs of \rf{app6} as its Mellin transform for $s\in [k_0+\oh,k_0+\frac{3}{2}]$.
 }
 where the  $R_0(x,-\sqL)$ given by \rf{jy1}. Unfortunately the subtlety of the equation is not 
 in $x$ and $s$ but in $u(T_0)$, since  $R(x,u)$ is the generating function of the Gelfand-Dikii polynomials.
 If the double scaling parameter $G$  is equal to zero this becomes a triviality and one indeed 
 finds $R(x,u)$ simply be an inverse Mellin transform of \rf{app6}.

\end{document}